\documentclass[prl,twocolumn]{revtex4}
\usepackage{graphicx}
\begin{document}
\title{Collisional decay of $^{87}$Rb Feshbach molecules at 1005.8 G}

\author{N. Syassen}
\affiliation{Max-Planck-Institut f{\"u}r Quantenoptik, Hans-Kopfermann-Stra{\ss}e 1, 85748 Garching, Germany}
\author{T. Volz}
\affiliation{Max-Planck-Institut f{\"u}r Quantenoptik, Hans-Kopfermann-Stra{\ss}e 1, 85748 Garching, Germany}
\author{S. Teichmann}
\altaffiliation{Present address: Centre for Atom Optics and Ultrafast Spectroscopy, Swinburne University of Technology, Hawthorn, VIC 3122, Australia}
\author{S. D{\"u}rr}
\affiliation{Max-Planck-Institut f{\"u}r Quantenoptik, Hans-Kopfermann-Stra{\ss}e 1, 85748 Garching, Germany}
\author{G. Rempe}
\affiliation{Max-Planck-Institut f{\"u}r Quantenoptik, Hans-Kopfermann-Stra{\ss}e 1, 85748 Garching, Germany}


\hyphenation{Fesh-bach}

\begin{abstract}
We present measurements of the loss-rate coefficients $K_{am}$ and $K_{mm}$ caused by inelastic atom-molecule and molecule-molecule collisions. A thermal cloud of atomic $^{87}$Rb is prepared in an optical dipole trap. A magnetic field is ramped across the Feshbach resonance at 1007.4~G. This associates atom pairs to molecules. A measurement of the molecule loss at 1005.8~G yields $K_{am}=2\times 10^{-10}$~cm$^3/$s. Additionally, the atoms can be removed with blast light. In this case, the measured molecule loss yields $K_{mm}=3\times10^{-10}$~cm$^3/$s.
\end{abstract}


\maketitle

Diatomic molecules associated from ultracold atomic gases using Feshbach resonances are in a highly-excited ro-vibrational state. An inelastic collision with another atom or molecule can lead to a vibrational de-excitation of the molecule. The difference in binding energy is released as kinetic energy in the relative motion of the molecule and the collision partner. This energy is typically much larger than the trap depth, so that both particles escape from the trap. The first experiments on association of molecules with Feshbach resonances \cite{regal:03,herbig:03,duerr:04,strecker:03,cubizolles:03,jochim:03a,xu:03,zwierlein:04} showed that the lifetimes of molecules made from bosonic atoms differ largely from the lifetimes of molecules made from fermionic atoms. Measurements in bosonic systems, $^{23}$Na and $^{133}$Cs, revealed loss-rate coefficients of typically $5\times10^{-11}$~cm$^3/$s \cite{mukaiyama:04,chin:05}. In fermionic systems, $^6$Li and $^{40}$K, the loss-rate coefficients far away from the Feshbach resonance are similar, but they can be suppressed by orders of magnitude when holding the magnetic field close to the Feshbach resonance \cite{cubizolles:03,regal:04a}. An explanation based on the Pauli exclusion principle for fermions was put forward \cite{petrov:04a,petrov:05}. Recent experiments showed that the short lifetimes in the bosonic species can be circumvented in an optical lattice \cite{thalhammer:06,volz:06}.

Here we present a measurement of the loss-rate coefficients $K_{am}$ and $K_{mm}$ for inelastic atom-molecule and molecule-molecule collisions, respectively. The molecules are associated from atomic $^{87}$Rb using the Feshbach resonance at 1007.4~G \cite{marte:02} with a width of $\Delta B= 0.2$~G \cite{volz:03,duerr:04a}. The loss measurements are performed at 1005.8~G. In one measurement, atoms and molecules are in the trap simultaneously. The loss in this measurement is dominated by inelastic atom-molecule collisions and reveals $K_{am}$. In another measurement, remaining atoms are removed from the trap using blast light \cite{xu:03,thalhammer:06} after associating the molecules. The loss in this measurement is dominated by inelastic molecule-molecule collisions and reveals $K_{mm}$. Both measurements are performed in thermal clouds. The value of $K_{mm}$ for a quantum degenerate cloud would be half as large \cite{burt:97}.

In our measurements, the molecules are associated from atoms in their absolute ground state. Spontaneous dissociation of Feshbach molecules into unbound atom pairs with lower-lying spin states as observed in $^{85}$Rb is therefore impossible \cite{thompson:05}. A previous measurement in $^{87}$Rb using photo-associated molecules set an upper limit of $K_{am}< 8\times10^{-11}$~cm$^3/$s \cite{wynar:00}. This limit is not applicable in the present experiment, because a different ro-vibrational state is investigated and the experiment is performed at a very different magnetic field.

A recent theoretical model for the state used here predicts $K_{am}=3\times10^{-10}$~cm$^3/$s at 1005.8~G \cite{smirne:cond-mat/0604183}. Previous models \cite{braaten:04,petrov:04} are only applicable if the magnetic field is less than $\Delta B$ away from the Feshbach resonance, which is not the case in the present experiment.

The experimental setup \cite{marte:02} is designed for the preparation of a BEC of $^{87}$Rb atoms in a magnetic trap. For the present experiment, the radio-frequency induced evaporation is stopped near the critical temperature $T_C$ of the phase transition to BEC. The atoms are then transferred into a crossed-beam optical dipole trap as described in Ref.~\cite{duerr:04a}. The dipole trap is operated at somewhat higher laser power than in Ref.~\cite{duerr:04a}. The measured trap frequencies are $(\omega_x,\omega_y,\omega_z) = 2\pi \times (95,154,200)$~Hz.

After transfer into the dipole trap, a magnetic field of $B=1007.6$~G is turned on rapidly and the spin of the atoms is transferred \cite{volz:03} to the hyperfine state $|F=1,m_F=1\rangle$. Next, $B$ is ramped across the Feshbach resonance at a rate of 0.4~G/ms. This associates atom pairs to molecules as described in Refs.~\cite{duerr:04,koehler:cond-mat/0601420}. As soon as molecules are forming during the ramp, they can undergo inelastic collisions. For molecules made from bosonic atoms, the inelastic collision rates are enhanced near the Feshbach resonance \cite{xu:03}. Hence, the molecule number can be maximized by jumping the magnetic field away from the resonance as fast as possible, once the molecules are created \cite{mark:05}. To this end, $B$ is jumped from 1007.35~G to 1005.8~G. We find experimentally that this combination of ramp speed and start point for the magnetic-field jump produces the maximum molecule number.

Immediately after jumping the field to 1005.8~G, remaining atoms can be removed from the trap by applying blast light for 0.3~ms as described in the appendix. This is followed by a variable hold time in the trap. This time is scanned in the loss measurements described below. At the end of this hold time, the trap is switched off. Right after release from the trap, the molecules are separated from remaining atoms using the Stern-Gerlach effect by applying a magnetic-field gradient of 120~G/cm for 1~ms. Immediately after this, the magnetic field is jumped to 1006.9~G and subsequently the molecules are dissociated into unbound atom pairs by ramping the magnetic field back across the Feshbach resonance to 1007.7~G at a rate of 0.8~G/ms. At the end of this ramp, the magnetic field is switched off rapidly. Finally, 7~ms after release from the trap, an absorption image is taken.

The molecule number decays as a function of hold time between association of the molecules and release from the trap. The loss of molecules from the trap can be described by the rate equation
\begin{eqnarray}
\label{eq-dndt}
\frac{d}{dt} n_m= -K_m n_m -K_{am} n_a n_m - K_{mm} n_m^2 \; ,
\end{eqnarray}
where $n_a$ and $n_m$ are the particle densities of atoms and molecules and $K_{am}$ and $K_{mm}$ are the loss-rate coefficients caused by inelastic atom-molecule and molecule-molecule collisions, respectively. $K_m$ represents molecule loss mechanisms which do not rely on collisions with other cold atoms or molecules. Such loss could be caused by background gas collisions, photo-dissociation by the dipole-trap light, or spontaneous decay into lower ro-vibrational levels. Our experimental results show that $K_m$ is negligible.

The loss of atoms during the hold time is also found to be negligible. This is because the atom number is either zero or much higher than the molecule number, so that inelastic atom-molecule collisions can only lead to loss of a small fraction of the atoms.

Volume integration of Eq.~(\ref{eq-dndt}) yields
\begin{eqnarray}
\label{eq-rate}
\frac{d}{dt} N_m= -K_m N_m -\frac{K_{am}}{V_{am}} N_a N_m - \frac{K_{mm}}{V_{mm}} N_m^2 \; ,
\end{eqnarray}
where $N_a$ and $N_m$ are the total number of atoms and molecules, respectively, and where we abbreviated
\begin{eqnarray}
\label{eq-volume}
\frac1{V_{im}} = \frac{1}{N_iN_m} \int n_in_m \; d^3x
\end{eqnarray}
for $i$ equal to $a$ or $m$ for atoms or molecules, respectively. $V_{im}$ is an effective volume and depends on the shape of the cloud, but not on the particle number. Note that $\langle n_m \rangle=N_m/V_{mm}$ is often referred to as the average density.

Assuming that $N_a$, $V_{am}$, and $V_{mm}$ are time independent, the rate equation (\ref{eq-rate}) can be integrated analytically using standard methods, yielding
\begin{eqnarray}
\label{eq-N(t)}
N_m(t) = \frac{N_0 \Gamma}{-N_0 \beta +(N_0\beta + \Gamma) e^{\Gamma t}} \; ,
\end{eqnarray}
where $\beta=K_{mm}/V_{mm}$ and $\Gamma=K_m + K_{am}N_a/V_{am}$ and $N_0$ is the molecule number at $t=0$.

\begin{figure}[tb]
\includegraphics[width=.45\textwidth]{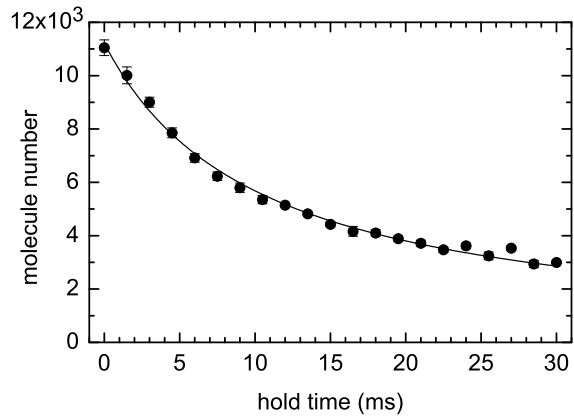}
\caption{\label{fig-Kmm}
Loss of molecules after blasting away the atoms. The solid line shows a fit of Eq.~(\ref{eq-N(t)}) to the data, yielding $K_{mm}$.
 }
\end{figure}

Fig.~\ref{fig-Kmm} shows experimental results for the molecule loss obtained after blasting away the atoms. In this measurement, $\Gamma=K_m$ because $N_a=0$. An unconstrained fit of Eq.~(\ref{eq-N(t)}) to the data yields a slightly negative value for $\Gamma$ which is unphysical. We therefore fix $\Gamma=0$ and obtain $\beta=(8\pm 1)\times10^{-3}/$s from the fit. The error bar is statistical.

\begin{figure}[tb]
\includegraphics[width=.45\textwidth]{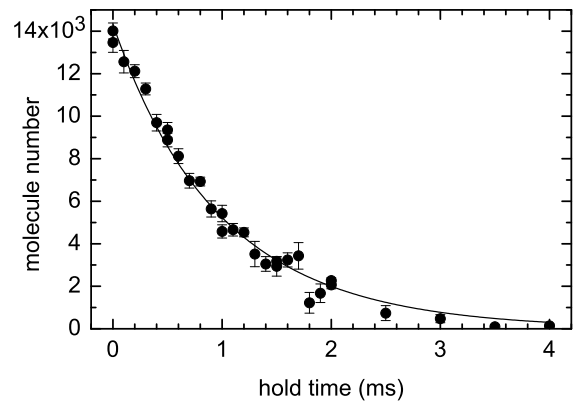}
\caption{\label{fig-Kam}
Loss of molecules in the presence of the atoms. The solid line shows a fit of Eq.~(\ref{eq-N(t)}) to the data, yielding $K_{am}$. Clearly, the loss here is much faster than in Fig.~\ref{fig-Kmm}.
 }
\end{figure}

Fig.~\ref{fig-Kam} shows experimental results without blasting away the atoms. Obviously, the presence of the atoms reduces the molecule lifetime substantially. An unconstrained fit of Eq.~(\ref{eq-N(t)}) to the data yields a slightly negative value for $\beta$ which is unphysical. We therefore fix $\beta$ to the value from Fig.~\ref{fig-Kmm} and obtain $\Gamma=(0.9\pm 0.1)/$ms from the fit. Again, the error bar is statistical.

In order to extract the loss-rate coefficients $K_{mm}$ and $K_{am}$, the effective volumes $V_{mm}$ and $V_{am}$ must be determined. This requires knowledge of the density distributions of the atomic and molecular cloud. This is a delicate issue, because the small cloud size makes direct measurements of the spatial distributions in the trap very hard. Theoretical modeling is also hard. Even the distributions at the beginning of the loss measurements are difficult to model, because the time evolution during molecule association is nontrivial. This is because, first, atomic pair correlations are crucial for a realistic treatment of the association process \cite{goral:04}. Second, the loss rate coefficients $K_{mm}$ and $K_{am}$ become relevant as soon as molecules start to form during the ramp. Third, these coefficients exhibit an unknown but probably strong magnetic-field dependence near the Feshbach resonance.

During the loss measurement, anharmonicities in the trap potential and elastic collisions between particles tend to randomize the motion leading towards a thermal distribution. Inelastic collisions, however, remove particles preferentially at the trap center, where the density is highest. The elastic scattering cross sections for the molecules are unknown. Hence, it is not clear which process dominates at what stage of the loss measurement. The evolution of the density distributions during the loss measurements is therefore a complex process with unknown parameters and unknown initial conditions.

In order to obtain an estimate for the effective volumes $V_{am}$ and $V_{mm}$, we assume that particles of the same species (atoms or molecules) are in thermal equilibrium. Our model does allow for a temperature difference between atoms and molecules. With this approximation, the spatial density distributions $n_a$ and $n_m$ are Gaussian and Eq.~(\ref{eq-volume}) yields the effective volume for species $i$
\begin{eqnarray}
V_{im}=(2\pi)^{3/2} \prod_{k=1}^3\sqrt{\sigma_{i,k}^2+\sigma_{m,k}^2} \; ,
\end{eqnarray}
where the index $k$ refers to the three directions in space and the one-dimensional (1D) root-mean-square (rms) radii of the Gaussians are $\sigma_{i,k}=\sqrt{k_BT_i/(m_i\omega_{i,k}^2)}$. The mass and temperature of species $i$ are labeled $m_i$ and $T_i$, respectively. The polarizability and the mass of a molecule are twice as large as for one atom, so that the trap frequencies for atoms and molecules are identical, i.e.\ $\omega_{a,k}=\omega_{m,k}$.

The temperature of the atoms and molecules is determined from time-of-flight measurements. For the atoms, the cloud size evolves as 
\begin{eqnarray}
\label{eq-a-exp}
\sigma_{a,k} (t)=\sqrt{\sigma_{a,k}^2(0)+\sigma_{v,a}^2 t^2} \; ,
\end{eqnarray}
where the trap is switched off at $t=0$ and $\sigma_{v,i}=\sqrt{k_BT_i/m_i}$ is the 1D rms-velocity of species $i$, which is independent of the spatial direction if the species is in thermal equilibrium. For the molecules, extra kinetic energy is added in the dissociation process \cite{mukaiyama:04,duerr:04a}, so that
\begin{eqnarray}
\label{eq-m-exp}
\sigma_{m,k} (t)=\sqrt{\sigma_{m,k}^2(0)+\sigma_{v,m}^2 t^2 +\sigma_{v,\rm dis}^2 t_{\rm rem}^2} \; .
\end{eqnarray}
Again, $t=0$ is chosen at the time of release from the trap. $\sigma_{v,\rm dis}$ reflects the extra kinetic energy released in the dissociation. $t_{\rm rem}$ is the remaining time of flight between dissociation and detection. In order to determine $T_m\propto\sigma_{v,m}^2$, we scan the time $t$ between release and detection in such a way that $t_{\rm rem}$ remains fixed. A fit of Eq.~(\ref{eq-m-exp}) is then equivalent to a fit of Eq.~(\ref{eq-a-exp}) with a modified value for $\sigma(0)$. Hence, the extracted temperature is insensitive to the dissociation heating.

The atomic cloud before molecule association typically contains $3.6\times10^5$ atoms at a temperature of $0.5~\mu$K very close to $T_C$. The cloud is almost purely thermal with only $6\times 10^3$ BEC atoms. The molecule association is accompanied by noticeable heating and substantial loss. The remaining atomic cloud contains $N_a=1.9 \times10^5$ atoms at a temperature of $T_a=1.0~\mu$K. There is no BEC in the remaining atomic cloud. The molecular cloud has a temperature of $T_m=1.5~\mu$K resulting in a peak phase-space density of $\sim 10^{-3}$ for the molecules. The center-of-mass motion of an atom pair is unchanged in the association and in a dilute thermal cloud the atomic pair correlations are uncorrelated from the center-of-mass motion of the pairs. Hence, our experiment should produce molecules with the same temperature as the initial atoms. This agrees reasonably with our measurements. The measured values of $N_a$, $T_a$, and $T_m$ vary by less than 10\% during the loss measurement. This justifies the assumption that they are time-independent, which was used to derive Eq.(\ref{eq-N(t)}).

The above values for $T_m$ and $T_a$ yield effective volumes of $V_{mm}=3.8\times 10^{-8}$~cm$^3$ and $V_{am}=4.6\times 10^{-8}$~cm$^3$. The resulting loss-rate coefficients are $K_{mm}=3\times 10^{-10}$~cm$^3/$s and $K_{am}=2\times 10^{-10}$~cm$^3/$s. Statistical errors on the rate coefficients are negligible compared to systematic errors.

The dominant systematic error arises from the problematic assumption that the clouds are in thermal equilibrium. For reasons discussed above, it is hard to measure or model the temporal evolution of the density distributions. On one hand, the association process preferentially populates the trap center, because the association is less efficient at low atomic density. On the other hand, inelastic collisions preferentially deplete the trap center. Both effects can lead to a misestimation of $V_{am}$ but the trends go into opposite directions. It is not clear which of the effects dominates. The resulting systematic error is hard to quantify. We speculate that a factor of 3 seems possible for $K_{am}$ as well as $K_{mm}$.

In conclusion, we measured the loss-rate coefficients $K_{am}$ and $K_{mm}$ for Feshbach molecules in $^{87}$Rb at 1005.8~G. These results yield valuable input for theoretical models of the three-atom and four-atom system. The measured value for $K_{am}$ agrees reasonably with the prediction $K_{am}=3\times10^{-10}$~cm$^3/$s of Ref.~\cite{smirne:cond-mat/0604183}.

\section*{Appendix: Blast light}
After associating molecules, remaining atoms can be pushed out of the trap using the radiation pressure of applied laser light \cite{xu:03,thalhammer:06}. The dipole trap used here is much deeper than the photon recoil energy, so that the blast light must drive a closed cycling transition, which is not possible when starting from the state $|F=1,m_F=1\rangle$. Hence, a first light field (the ``pump light") optically pumps the atoms from $|F=1,m_F=1\rangle$ to $|F=2,m_F=2\rangle$. A second light field (the ``cycling light") then drives a cycling transition. We only drive transitions on the $5\,^2\!S_{1/2}\leftrightarrow 5\,^2\!P_{3/2}$ resonance line at $780$~nm. 

As the blast light is applied while the atoms are in a magnetic field of 1005.8~G, some effort is needed to obtain light stabilized at the required frequencies. Two frequency-stabilized lasers are already in use in the experiment for operating the magneto-optical trap (MOT): one is locked close to the MOT transition $|F=2\rangle \leftrightarrow |F'=3\rangle$ and another laser is locked to the repump transition $|F=1\rangle \leftrightarrow |F'=2\rangle$. Quantum numbers with a prime refer to excited states.

The cycling laser resonantly drives the $\sigma^+$ transition $|F=2,m_F=2\rangle \leftrightarrow |m_I'=3/2,m_F'=3\rangle$ at 1005.8~G, which is 1405~MHz blue detuned from the $B=0$ MOT transition. Note that at 1005.8~G, the relevant ground states are characterized by good quantum numbers $F,m_F$, while the excited states are characterized by good quantum numbers $m_I',m_F'$. A beat lock stabilizes the frequency of the cycling laser relative to the MOT laser.

In order to deplete the state $|F=1,m_F=1\rangle$, we use pump light that is 62~MHz red detuned from the $B=0$ repump transition. The light is obtained from the repump laser using an acousto-optical modulator (AOM) and resonantly drives the $\pi$ transition $|F=1,m_F=1\rangle \leftrightarrow |m_I'=3/2,m_F'=1\rangle$ at 1005.8~G.

From the excited state populated by this pump light, atoms can decay back to the initial state, into the desired state $|F=2,m_F=2\rangle$  or into the undesired state $|F=2,m_F=1\rangle$. The experiment shows, that the branching ratio for decay into the undesired state is only a few percent. Still, we use a third light field to deplete this state. We use another AOM to obtain light that is 87~MHz red detuned from the $B=0$ MOT transition. This light resonantly drives the $\sigma^+$ transition $|F=2,m_F=1\rangle \leftrightarrow |m_I'=3/2,m_F'=2\rangle$ at 1005.8~G.

Note that the excited state $|m_I'=3/2,m_F'=1\rangle$ cannot decay into ground states with $m_F=0$, because at 1005.8~G the quantum number $m_I=3/2$ is conserved during the decay and the $J=1/2$ ground state does not have sub-states with $m_J=-3/2$.

All three blast fields are operated a factor of ten or more above saturation intensity. They are on simultaneously. The trap depth is a few microkelvin, so that approximately five directed photon recoil momenta should add sufficient kinetic energy for an atom to leave the trap. This momentum should be accumulated after $0.2~\mu$s. The estimated acceleration is $10^5$~m/s$^2$, so that the cloud radius of $10~\mu$m is estimated to be traversed in $15~\mu$s. But the experiment shows that the blast light needs to be on for $300~\mu$s, in order to remove all atoms from the trap. The origin of this discrepancy is unclear.

Application of the three blast beams reduces the molecule number by approximately 30\%. A recent experiment, performed after the measurements reported here, demonstrated that atoms can be removed by applying a microwave field and only the cycling laser \cite{thalhammer:06}. This has the advantage that the loss of molecules during application of the blast light becomes negligible \cite{volz:06}.

\end{document}